\documentclass[journal,comsoc]{IEEEtran}
\usepackage{cite}
\usepackage{graphicx}
\usepackage{amsmath}
\usepackage{times}
\usepackage{latexsym}
\usepackage{graphicx}
\usepackage{bm}
\usepackage{amssymb}
\usepackage[center]{caption2}
\usepackage{stfloats}
\usepackage{cases}
\usepackage{subfigure}
\usepackage{array}
\usepackage{setspace}
\usepackage{fancyhdr}
\usepackage{citesort}
\usepackage{colortbl}
\usepackage{CJK}
\usepackage{tikz}
\usepackage{float}
\usepackage{color}
\usepackage{siunitx}
\usepackage{bm}
\usepackage{mathrsfs}
\usepackage{diagbox}

\renewcommand{\QED}{\QEDopen}

\allowdisplaybreaks[4]

\renewcommand{\baselinestretch}{0.97}

\newcaptionstyle{mystyle2}{%
\captionlabel $.$ \, \captiontext \par} \captionstyle{mystyle2}
\renewcommand\thesubsection{\Alph{subsection}}

\newcommand{\tabincell}[2]{\begin{tabular}{@{}#1@{}}#2\end{tabular}}  

\begin{document}
\title{Deep Reinforcement Learning for Joint Beamwidth and Power Optimization in mmWave Systems}
\author{\authorblockN{Jiabao Gao, Caijun Zhong, Xiaoming Chen, Hai Lin and Zhaoyang Zhang}
\thanks{J. Gao, C. Zhong, X. Chen, and Z. Zhang are with the College of Information Science and Electronic Engineering, Zhejiang University, Hangzhou, China, and also with the Zhejiang Provincial Key Laboratory of Information Processing, Communication and Networking, Hangzhou. (email: caijunzhong@zju.edu.cn).}
\thanks{H. Lin is with the Department of Electrical and Information Systems, Osaka Prefecture University, Osaka 599-8531, Japan (email: hai.lin@ieee.org).}}
\maketitle

\begin{abstract}
This paper studies the joint beamwidth and transmit power optimization problem in millimeter wave communication systems. A deep reinforcement learning based approach is proposed. Specifically, a customized deep Q network is trained offline, which is able to make real-time decisions when deployed online. Simulation results show that the proposed approach significantly outperforms conventional approaches in terms of both performance and complexity. Besides, strong generalization ability to different system parameters is also demonstrated, which further enhances the practicality of the proposed approach.
\end{abstract}

\begin{keywords}
Millimeter wave, beamwidth selection, power allocation, joint optimization, deep reinforcement learning
\end{keywords}

\section{INTRODUCTION}
Due to the abundant spectrum at the millimeter frequency band, millimeter wave (mmWave) communication \cite{mmWave} has been visioned as a key enabling technology to overcome the spectrum shortage challenge in next generation communication systems. With high operating frequency, mmWave communications suffer from severe path loss and sensitivity to blockage\cite{mmChannel}. Fortunately, highly directional beams can be formed with a large number of antennas, which can effectively alleviate above issues\cite{pencil}.

To fully unleash the potential of mmWave communications, fine alignment of the transmitting and receiving beams is of crucial importance, which gives rise to the intrinsic alignment gain and data transmission time trade-off. In addition, for the scenario with multiple transmitter-receiver pairs, interference management is essential to further improve the system performance, which requires sophisticated design of beamwidth and power control mechanism.

Thus far, several works have proposed different approaches to address the above issues. In work \cite{Baseline}, two suboptimal algorithms were proposed, namely underestimation and overestimation of the interference. The former neglects interference and decomposes the joint optimization problem into several single pair subproblems which is convex and can be easily solved. The latter overestimates interference and only activates part of pairs without severe mutual interference. Besides, work \cite{review1} studied the transmit power control problem in uplink mmWave cellular networks, while a heuristic algorithm based on simulated annealing was proposed to jointly optimize the power and beamwidth in work \cite{heuristic}. Due to the various simplifying assumptions adopted in the aforementioned works, the proposed approaches in general yield suboptimal solutions.

Responding to this, we aim to tackle the joint beamwidth and power control problem. Specifically, motivated by its great potential of handling  complex non-convex problems in communications \cite{ML1,ML2,ML3}, we pursue a artificial intelligence (AI) based design here. However, the most popular supervised learning paradigm is not suitable, mainly due to the prohibitive cost of labeling. Therefore, we propose to use deep reinforcement learning (DRL), a mechanism which does not require labels naturally. Recently, \cite{R1} also proposed to use deep Q network (DQN) to address decision making problems in beam optimization. However, the current work differs from \cite{R1} in terms of both objective function and action design. Besides, \cite{R1} only considers beam selections, while this work jointly optimizes the transmit power and beamwidth to further enhance system performance. In particular, a customized DQN is designed to solve the decision making problem. We carefully preprocess the channel state information (CSI) and noise power density to formulate the state tuple, which can enhance effective learning of DRL framework. After offline training, the DQN is subsequently deployed online for real-time decision of the beamwidths and transmit power. Extensive simulation results are provided to evaluate the performance of proposed algorithm, which demonstrates the superior performance of the proposed method over conventional approaches.

The rest of this paper is organized as follows. Section II introduces the system model and formulates the optimization problem. Section III presents the detailed design of the proposed DRL based approach, and its superiority is validated by extensive simulation results in Section IV. Finally, the paper is concluded in Section VI.

\section{System model and Problem formulation}
\label{System model and problem formulation}
\begin{figure}[htbp]
\centering
\includegraphics[width=0.4\textwidth]{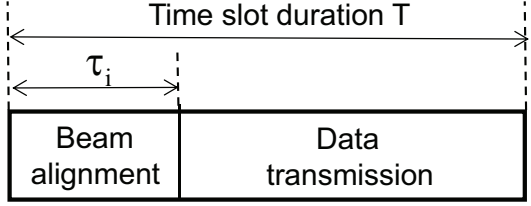}
\caption{Time slot segmentation of link $i$. T denotes the time slot duration and $\tau_i$ denotes beam alignment time.}
\label{timeslot}
\end{figure}

We consider a mmWave system consisting of $N$ transmitter-receiver pairs. Each time slot can be divided into two phases, namely beam alignment and data transmission, as shown in Fig. \ref{timeslot}. In the beam alignment phase\footnote{Beam alignment can be further divided into sector-level and beam-level alignment. In this paper, we assume that sector-level alignment has already been established as in \cite{SectorLevel}.}, each pair decides the optimal transmitting and receiving beam directions that maximize the signal-to-noise ratio (SNR) within their sectors by searching over all possible combinations, as specified in IEEE 802.15.3c\cite{Baseline}. In the data transmission phase, data is transmitted and received using the selected beams.

Denote $\psi_i^t$ and $\psi_i^r$ as sector-level beamwidths, $\varphi_i^t$ and $\varphi_i^r$ as beam-level beamwidths at the transmitter and receiver of link $i$ respectively. Then, the number of possible combinations is $\lceil \frac{\psi_i^t}{\varphi_i^t} \rceil \lceil \frac{\psi_i^r}{\varphi_i^r} \rceil$. Denote $T_p$ as pilot transmission time of each combination, then the total alignment time is
\begin{equation}
\tau_i(\varphi_i^t,\varphi_i^r)=\lceil \frac{\psi_i^t}{\varphi_i^t} \rceil \lceil \frac{\psi_i^r}{\varphi_i^r} \rceil T_p.
\end{equation}

\begin{figure}[htbp]
\centering
\includegraphics[width=0.4\textwidth]{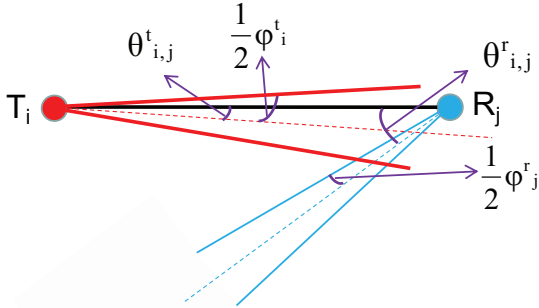}
\caption{The black solid line denotes the boresight between transmitter $i$ and receiver $j$, and the dotted lines denote the angle bisectors of transmitting and receiving beams.}
\label{angle}
\end{figure}

We adopt the antenna model presented in \cite{directionalgain}, then the transmission and reception gains at transmitter $i$ and receiver $j$ toward each other can be expressed as
\begin{equation}
\label{TransmitGain}
g_{i,j}^t(\theta_{i,j}^t,\varphi_i^t)=
\left\{
             \begin{array}{ll}
             \frac{2\pi-(2\pi-\varphi_i^t)z}{\varphi_i^t}, & \text{if} \,\, |\theta_{i,j}^t|\le\frac{\varphi_i^t}{2}\\
             z, & \text{otherwise}
             \end{array}
\right.,
\end{equation}
\begin{equation}
\label{ReceiveGain}
g_{i,j}^r(\theta_{i,j}^r,\varphi_j^r)=
\left\{
             \begin{array}{ll}
             \frac{2\pi-(2\pi-\varphi_j^r)z}{\varphi_j^r}, & \text{if} \,\, |\theta_{i,j}^r|\le\frac{\varphi_j^r}{2}\\
             z, & \text{otherwise}
             \end{array}
\right.,
\end{equation}
where $\theta_{i,j}^t$ and $\theta_{i,j}^r$ are the angles between the boresight of transmitter $i$ and receiver $j$ and the angle bisectors of transmitting and receiving beams, $z$ is the side lobe gain, as illustrated in Fig .\ref{angle}.

Denote $g_{i,j}^c$ as the channel gain between transmitter $i$ and receiver $j$, $p_i$ as the transmission power of transmiter $i$, $N_0$ as the thermal noise power spectral density and $W$ as the system bandwidth. Then, the signal-to-interference-pluse-noise ratio (SINR) of the $i$-th link can be given as
\begin{equation}
\label{SINR}
\text{SINR}_i=\frac{p_ig_{i,i}^tg_{i,i}^rg_{i,i}^c}{\sum \limits_{k=1,k\neq i}^{N}p_kg_{k,i}^tg_{k,i}^rg_{k,i}^c+WN_0}.
\end{equation}

Therefore, the joint beamwidth selection and power allocation optimization problem can be formulated as
\begin{align}
\text{P1}: \qquad
&\underset{\bm{\varphi}^t,\bm{\varphi}^r,\bm{p}}{\text{max}}\quad W \sum \limits_{i=1}^N(1-\frac{\tau_i}{T})\text{log}_2({1+\text{SINR}_i)}\\
& \begin{array}{r@{\quad}l@{\quad}l}
\, \text{s.t.} &\varphi_i^t \le \psi_i^t,  &i=1,2\ldots,N,  \\
&\varphi_i^r \le \psi_i^r,  &i=1,2\ldots,N, \\
&\psi_i^t\psi_j^rT_p/T \le \varphi_i^t \varphi_j^r,   &i,j=1,2\ldots,N, \\
&0 \le p_i \le p^{\text{max}},  &i=1,2\ldots,N, \\
\end{array}
\end{align}
where the first two constraints ensure the beamwidths in beam-level is strictly smaller than that in sector-level, the third constraint means that the beam alignment time can not exceed the entire time slot duration, and the fourth constraint specifies the maximal transmission power limit $p^{\text{max}}$. Since this problem is non-convex, it is challenging to solve by conventional optimization approaches.

\section{Deep Reinforcement learning based approach}
In this Section, we propose a DQN based approach to solve the above problem, and we start with a brief introduction of the DQN algorithm.

\subsection{Brief introduction to DQN}
\begin{figure}[htbp]
\centering
\includegraphics[width=0.4\textwidth]{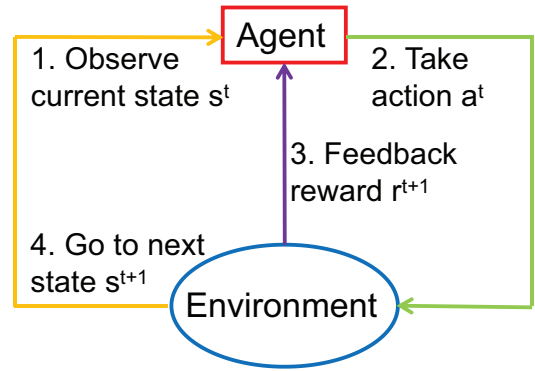}
\caption{The agent-environment interaction in MDP.}
\label{rl}
\end{figure}

Fig. \ref{rl} illustrates the typical agent-environment interaction in a Markov Decision Process (MDP). At time step $t$, the agent takes an action $a^t\in A$ by observing current state $s^t\in S$ to interacts with the environment, where $S$ and $A$ are the sets of states and actions, respectively. One time step later, as the consequence of its action, the agent receives a reward $r^{t+1}$ and moves into a new state $s^{t+1}$. The goal of reinforcement learning (RL) is to maximize the long-term rewards\cite{RL}. Specifically, it aims to learn the policy that yields maximal cumulative discounted reward function as follows
\begin{equation}
G^t = \sum \limits_{k=0}^{\infty}\gamma^kr^{t+k+1},
\label{reward}
\end{equation}
where $0\le \gamma \le1$ is called the discount rate to discount rewards of later time slots.

Q learning, as one of the most popular RL algorithms, maintains a Q table to record Q values of all (state, action) pairs. Under policy $\pi$, the Q function of agent with action $a$ in state $s$ is defined as
\begin{equation}
\label{Qfunc}
Q_\pi(s,a) = \mathbb{E}[G^t|s^t=s,a^t=a;\pi],
\end{equation}
where the expectation is taken with respect to the environment and policy.

Through learning from trajectories, the Q table is updated to approach the real table under optimal policy $\pi^*$, which is achieved when the following Bellman optimality equation is satisfied\cite{RL}
\begin{equation}
\label{bellman}
Q_{\pi^*}(s^t,a^t) = \underset{a^{t+1}}{\text{max}}(\mathbb{E}[r^{t+1}+\gamma Q_{\pi^*}(s^{t+1},a^{t+1})]).
\end{equation}

To further tackle the curse of dimensionality with continuous state variables, DQN\cite{DQN} is proposed to directly predict Q value for any (state, action) pair by a deep neural network. Then, the policy is parameterized by the weights of the Q network $\bm{\theta}$, which can be updated as\footnote{{As will be explained later, $\gamma$ is set to 0, so some critical parts of DQN like target network is not involved.}}
\begin{equation}
\bm{\theta}^{t+1}=\bm{\theta}^{t}\\
+\eta(y^t-Q(s^t,a^t;\bm{\theta}^t))\nabla{Q(s^t,a^t;\bm{\theta}^t)},
\end{equation}
where $y^t=r^{t+1}+\gamma \underset{a^{t+1}}{\text{max}}Q(s^{t+1},a^{t+1};\bm{\theta}^{t})$ is the optimal Q value, $\eta$ is the learning rate, and $\nabla$ denotes the derivative operator.

\subsection{Customized DQN design}
In this subsection, we present the details about the customized DQN design for P1, including state, action, reward, network architecture and training strategy.
\subsubsection{State}
As in \cite{Baseline}, we assume that perfect CSI is available. The state is the observation that agent can get from the environment. In the current problem, the observation of a particular link contains the channel gain of the link, $N-1$ interfering channel gains from this link's transmitter to other links' receivers (ITO), $N-1$ interfering channel gains from other links' transmitters to this link's receiver (IFO), and noise power. Intuitively, one simple way is to directly use the raw data as state. However, it turns out that the performance is not satisfactory, which implies that proper preprocessing is of critical importance. Motivated by this, we first normalize all elements by the channel gain of the link to better expose the characteristics of relative interference and noise level. Then, we choose dB as unit to further reduce the variance of input and facilitate efficient training. Therefore, the state tuple of link $i$ at certain time slot contains totally $2N-1$ elements, which can be expressed as
\begin{equation}
\bm{s}_i = \bm{s}_i^{\text{ITO}} \cup \bm{s}_i^{\text{IFO}} \cup \bm{s}_i^{\text{Noise}}, \quad i=1,2,\cdots,N,
\end{equation}
where $\bm{s}_i^{\text{ITO}}=\{g_{i,1}^c,\ldots,g_{i,i-1}^c,g_{i,i+1}^c,\ldots,g_{i,N}^c\}/g_{i,i}^c$, $\bm{s}_i^{\text{IFO}}=\{g_{1,i}^c,\ldots,g_{i-1,i}^c,g_{i+1,i}^c,\ldots,g_{N,i}^c\}/g_{i,i}^c$ and $\bm{s}_i^\text{Noise}=\{WN_0\}/g_{i,i}^c$.

\subsubsection{Action}
To achieve the optimal performance, the actions at all links should be jointly optimized. In such case, the action number will grow exponentially with the number of links, making effective learning impossible. Therefore, to avoid such problem, we let the DQN make decision for every link separately. For each link, the action is the combination of beamwidths for transmitter and receiver and transmit power. Assume transmitter and receiver use the same beamwidth as in \cite{Baseline}, i.e., $\varphi^t = \varphi^r = \varphi$, then the action of link $i$ at certain time step can be expressed as $\bm{a}_i=[p_i,\varphi_i]$.

\subsubsection{Reward}
Since the objective is to maximize the instantaneous effective sum rate, the discount rate $\gamma$ should be set to 0. Therefore, $G^t$ in Equation \ref{reward} reduces to $r^{t+1}$, where $r^{t+1}$ is the effective sum rate at time slot $t+1$.

\subsubsection{Q Network architecture}
The neural network architecture for Q value estimation is designed by cross validation. In our experiments, a fully-connected network consisting of two hidden layers with 128 and 64 neurons and an output layer with the number of neurons equivalent to total action number works well.

\subsubsection{DQN training}
The initial learning rate of neural network is 0.001, batch size is 256 and weights are updated by the Adam optimizer. First, we only generate data for 2000 episodes to fill the replay buffer. Then, in order to balance exploration and exploitation during training, $\epsilon-\text{greedy}$ policy is adopted\cite{RL}. Specifically, the DQN will randomly choose an action from the action set with a small probability $\epsilon$, rather than always choose action with the maximal Q value. The initial $\epsilon$ is set to 0.2, and gradually decreases to 0 in 100000 episodes. After that, we continue to train with $\epsilon=0$ for 10000 episodes. Notice that, in practice, a large amount of training data can be generated automatically based on the model. However, mismatch may exist between the assumed channel model and the actual propagation environment. In such case, online fine-tuning can be adopted to compensate the model mismatch, by continuing training with real environmental data.

\section{Simulation Results}
\label{simulation}
In this section, extensive simulation results are provided to demonstrate the performance of the proposed DRL based approach. The random selection algorithm and underestimation of interference algorithm are used as benchmark for performance comparison. We assume that all the transmitter-receiver pairs are distributed randomly in a square area with a side length of $L$ m. As in \cite{mmChannel}, the following mmWave channel pathloss model is used:
\begin{equation}
\label{pathloss}
l(R)=\mathbb{I}(p(R))l_{\text{LoS}}(R)+\mathbb{I}(1-p(R))l_{\text{NLoS}}(R),
\end{equation}
where $l_{\text{LoS}}(R)=C_{\text{LoS}}R^{-\alpha_{\text{LoS}}}$ and $l_{\text{NLoS}}(R)=C_{\text{NLoS}}R^{-\alpha_{\text{NLoS}}}$ account for the line-of-sight (LoS) and non-line-of-sight (NLoS) loss, respectively. Also, $\mathbb{I}(x)$ is the indicator function which returns 1 when $x=1$ and 0 otherwise, while $p(R)$ is a boolean random variable with probability $e^{\beta R}$ being 1, and $R$ denotes distance between the transmitter and receiver. The following set of parameters are used in simulation as in\cite{mmChannel}.
\begin{table}[htbp]
\begin{tabular}{|c|c|}
\hline
Parameter name & Value \\ \hline
Carrier frequency & 28 GHz \\ \hline
System bandwidth & 1 GHz \\ \hline
Reference distance & 5 m \\ \hline
LoS path loss & $\alpha_{\text{LoS}}=2$, $C_{\text{LoS}}=-60\text{dB}$ \\ \hline
NLoS path loss & $\alpha_{\text{NLoS}}=4$, $C_{\text{NLoS}}=-70\text{dB}$   \\ \hline
Blockage model & $\beta=0.006$ \\ \hline
NLoS small-scale fading & Nakagami fading of parameter 3 \\ \hline
Noise power density &  $N_0=-145\text{dBm/Hz}$ \\ \hline
Sector-level beamwidth & $\psi_i^t=\psi_i^r=90^\circ$ for all $i$ \\ \hline
Side lobe gain & $z=0.1$ \\ \hline
Pilot transmission time & $T_p/T=0.001$ \\ \hline
\end{tabular}
\centering
\caption{Simulation parameters}
\label{parameters}
\end{table}

Next, we investigate the impact of several key factors and parameters, including action discretization, network density and area size. All the tables and curves about testing performance are obtained by averaging over 500 independent experiments.

\subsection{Impact of action discretization}
\begin{figure}[htbp]
\centering
\includegraphics[width=0.50\textwidth]{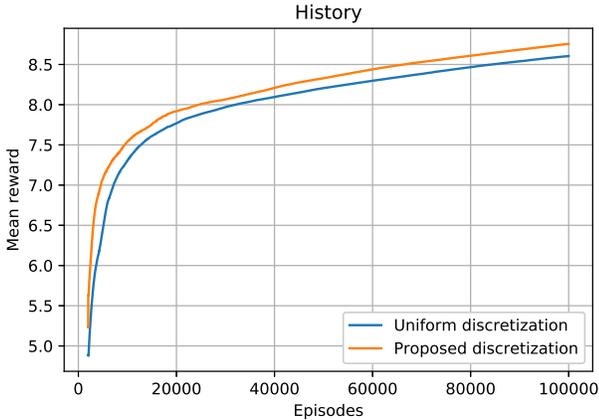}
\caption{Training histories with different discretization patterns when $L=20$, $N=10$. The vertical coordinate refers to the mean reward from the first episode to the current episode, in order to smooth the curve and highlight the growth trend.}
\label{history}
\end{figure}

For DQN training, the actions are discretized depending on how they affect the system performance. According to equations (1-4), the SINR is a linear function with respect to the transmit power and the reciprocal of the beamwidth. Therefore, we propose to uniformly discretize the transmit power and the reciprocal of the square of beamwidth in $[P_{min},P_{max}]$ and $[1/\varphi_{max}^2,1/\varphi_{min}^2]$, respectively, where $P_{min}$ and $P_{max}$ are the minimal and maximal transmit power, and $\varphi_{min}$ and $\varphi_{max}$ are the minimal and maximal beamwidth. In simulation, we set $P_{min} = \text{2 dBm},P_{max} = \text{30 dBm},\varphi_{min}=\ang{3},\varphi_{max}=\ang{30}$. Besides, we use only 8 values for both the transmit power and beamwidth set to balance the training complexity and performance. To illustrate the performance of the proposed approach, the heuristic uniform approach is used as a benchmark, where both the transmit power and bemwidth are uniformly discretized. As can be observed in Fig. \ref{history}, the proposed scheme achieves better performance compared with the uniform scheme.

To demonstrate the superiority of the proposed approach, it is necessary to see the performance gap when compared with the performance of exhaustive search (ES) scheme. Due to the complexity of the ES scheme, it is only feasible to make the comparison in relatively small scale systems. As shown in Table \ref{ES}, the performance of the DQN approach is close to that of the ES scheme, and is far superior to that of the random scheme.

\begin{table}[htbp]
\begin{tabular}{|c|c|c|c|c|}
\hline
$N$ & \tabincell{c}{ES \\ (G bits/slot)} & \tabincell{c}{Percentage \\ of DQN (\%)} & \tabincell{c}{Percentage \\ of Random (\%)} \\ \hline
2 & 19.58 & 98.72 & 40.42 \\ \hline
3 & 28.74 & 97.56 & 38.65 \\ \hline
4 & 35.18 & 96.09 & 39.35 \\ \hline
5 & 45.28 & 95.02 & 39.80\\ \hline
\end{tabular}
\centering
\caption{Network throughput performance of different approaches when $L=20$. Transmit power and beamwidth are discretized the same way as above with 4 optional values each.}
\label{ES}
\end{table}

\subsection{Impact of network density}
\begin{figure}[htbp]
\centering
\includegraphics[width=0.50\textwidth]{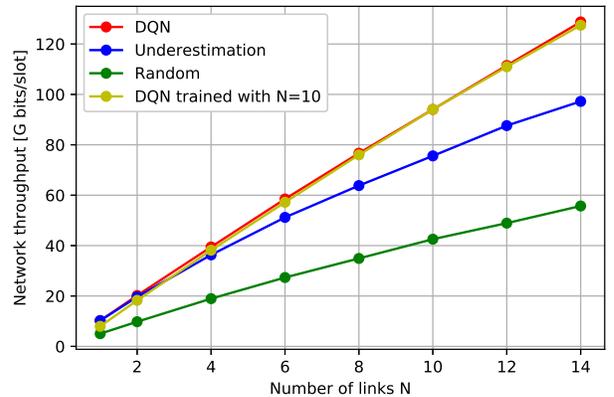}
\caption{The impact of network density on throughput.}
\label{throughput}
\end{figure}

Fig. \ref{throughput} illustrates the impact of network density on the network throughput with $L=20$ and different $N$. It can be observed that the throughput increases with $N$, while the slope gradually decreases due to accumulated mutual interference. Also, the proposed DQN based approach consistently outperforms the underestimation of interference baseline\cite{Baseline} when $N \ge 4$, and the performance gap becomes larger as $N$ increases. The reason is that the proposed DQN method takes into account of the interference during the design, hence can achieve superior performance in crowded scenarios with severe interference.

In practice, the network density may change over time. Therefore, it is of paramount importance to investigate the generalization ability of the proposed approach, which is a also key challenge when implementing AI based wireless communication systems \cite{Challenges}. With a well trained network with $N$, if the actual link number $N^*$ is smaller than $N$, we simply consider the extra links as virtual links and pad each real link's state tuple with $N^*-N$ zeros. On the other hand, if $N^*$ is greater than $N$, we sort the interfering channel gains of each link and only keep the largest $N^*-1$ ones. As illustrated in Fig. \ref{throughput}, where the red curve corresponds to the case with customized training for different $N$, while the yellow curve is trained for $N=10$, we observe that the two curves almost overlap for different $N$, which demonstrates the superior generalization capability of the proposed DQN.

\subsection{Impact of area size}
\begin{figure}[htbp]
\centering
\includegraphics[width=0.50\textwidth]{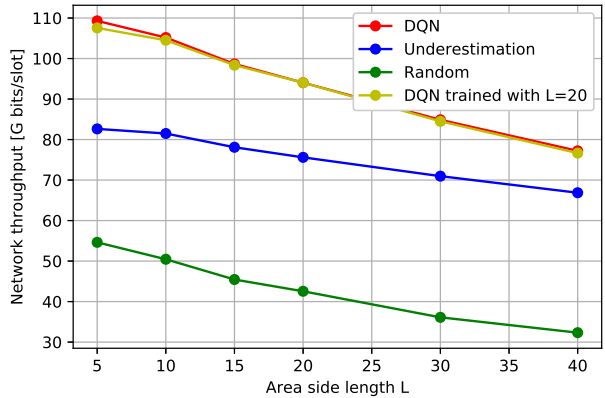}
\caption{The impact of area side length on throughput.}
\label{area}
\end{figure}

Fig. \ref{area} illustrates the impact of area side length on network throughput with $N=10$ and different $L$. As can be readily observed, the throughput decreases with $L$, which is straightforward due to increased pathloss. Also, the proposed DQN based approach consistently outperforms the underestimation of interference baseline under various $L$, while the performance gap decreases with larger $L$ since the impact of interference becomes insignificant. The generalization ability on $L$ is also investigated. As we can see, the network trained with $L=20$ works well for a wide range of $L$ with very little performance deterioration.

\subsection{Complexity}
When $L=20$, $N=10$, the typical offline training time of the proposed approach on a GeForce GTX 1080 Ti GPU is about one hour. However, the online testing phase only needs to execute simple forward computation, which is much faster than the gradient descent process involved in the underestimation of interference baseline approach. When $N=4$ and $N=10$, the average running time of the baseline approach is 207.40 and 513.04 ms, while the DQN based approach takes 0.98 and 0.99 ms, respectively, which is hundreds of times faster.

\section{CONCLUSION and future work}
\label{conclusion}
In this paper, we have proposed a DRL based approach to solve the joint beamwidth and power allocation problem in mmWave communication systems. A customized DQN is designed, and heuristic tricks are used to tackle the generalization issue. Simulation results show that the proposed approach significantly outperforms the conventional suboptimal approaches in terms of both performance and complexity. Besides, the proposed DQN has strong generalization ability, which makes it extremely desirable for practical implementation. In the future, we will consider the use of advanced DRL algorithms such as deep deterministic policy gradient\cite{DDPG} to optimize in continuous action domain.

\end{document}